\documentclass{PoS}
\usepackage{subfigure}
\pdfoutput=1	

\title{Performance studies of the new stereoscopic Sum-Trigger-II of MAGIC after one year of operation}

\ShortTitle{Sum-Trigger-II performances}

\author{\speaker{Francesco Dazzi} \\ Max-Planck-Institut f\"{u}r Physik, M\"{u}nchen (Germany) \\ E-mail: \email{dazzi@mppmu.mpg.de}}
\author{Diego Herranz Lazaro \\ Universidad Complutense, Madrid (Spain) \\ E-mail: \email{diegoherranz@fis.ucm.es}}
\author{Marcos L\'{o}pez \\ Universidad Complutense, Madrid (Spain) \\ E-mail: \email{marcos@sagan.gae.ucm.es}}
\author{Daisuke Nakajima \\ Institute Cosmic Ray Research, Tokyo (Japan) \\ E-mail: \email{dnakajim@icrr.u-tokyo.ac.jp}}
\author{Jezabel Rodriguez Garcia \\ Max-Planck-Institut f\"{u}r Physik, M\"{u}nchen (Germany) \\ E-mail: \email{jezabel.rg@gmail.com}}
\author{Thomas Schweizer \\ Max-Planck-Institut f\"{u}r Physik, M\"{u}nchen (Germany) \\ E-mail: \email{tschweiz@mppmu.mpg.de}}

\abstract{MAGIC is a stereoscopic system of two Imaging Air Cherenkov Telescopes (IACTs) located at La Palma (Canary Islands, Spain) and working in the field of very high energy gamma-ray astronomy. It makes use of a traditional digital trigger with an energy threshold of around 55\,GeV. A novel trigger strategy, based on the analogue sum of signals from partially overlapped patches of pixels, leads to a lower threshold. In 2008, this principle was proven by the detection of the Crab Pulsar at 25\,GeV by MAGIC in single telescope operation. During Winter 2013/14, a new system, based on this concept, was implemented for stereoscopic observations after several years of development. In this contribution the strategy of the operative stereoscopic trigger system, as well as the first performance studies, are presented. Finally, some possible future improvements to further reduce the energy threshold of this trigger are addressed.}

\FullConference{The 34th International Cosmic Ray Conference, \\ 30 July- 6 August, 2015 \\ The Hague, The Netherlands}

\begin{document}

\section{Introduction}
	\label{sec:introduction}
The MAGIC (Major Atmospheric Gamma Imaging Cherenkov) telescopes belong to the $3^{rd}$ generation of ground-based instruments that operate in the Very High Energy domain of gamma-ray Astronomy, from few tens of GeV to few tens of TeV. They make use of the well established Imaging Air Cherenkov Technique (IACT) \cite{iact} for reconstructing the energy and direction of gamma rays from the Universe. The telescopes have a large reflective surface that focuses the Cherenkov light produced in the atmosphere by a gamma ray into a camera equipped with photo-sensors. The light, that forms in the camera plane an image of the extended air shower, is converted in electrical signals processed by dedicated electronics. The electronics chain is split in two branches. In the first one the signal is continuously sampled thank to a readout system based on DRS4 chip, in the second one the signal is processed by a digital trigger system that selects the events of interest and informs the readout to store them into a mass storage \cite{lt1_trigger}. Since Winter 2013/14, the MAGIC telescopes have been equipped with an additional trigger system, dubbed Sum-Trigger-II. Usually, Sum-Trigger-II is activated for special physics cases that need a low energy threshold, as for instance Pulsars.

\section{System description}
	\label{sec:system_description}
The IACT telescopes aim to record the Cherenkov light emitted by air showers initiated by gamma rays, messengers that bring information about the astrophysical processes \cite{gamma_astronomy}. The high background rate originated by the Light of the Night Sky (LoNS) and showers not initiated by gamma rays makes this goal really challenging. It is common to adopt fast trigger strategies that reject at the hardware level most of the background due to the LoNS, so the DAQ (Data AcQuisition) load is kept at a sustainable rate. The further discrimination between events initiated by gamma rays and other primary particles is done off-line through sophisticated software algorithms that parametrize the shower's morphology. A typical digital trigger approach is to first digitize the photo-sensor pulses that exceed a programmable threshold, secondly to apply the proper delay to cancel the skew artificially introduced by the detector, and finally to search for coincidences of \emph{n} neighbour pixels within a narrow time window (trigger gate) of few nanoseconds. It is effective to keep the trigger gate quite small because this increases the background rejection power, but it cannot be arbitrarily reduced. The electronics performances and the intrinsic time spread of gamma events fix the lower limit that needs to be analysed case by case. An alternative trigger strategy, more effective for low energy events, is based on the pile-up of analogue pulses from neighbouring Photo-Multipliers Tubes (PMTs). The discriminator threshold is applied only after the summing stage, therefore also the PMTs with a small charge contribute to the final trigger decision.

\subsection{Trigger strategy for low energy events}
	\label{subsec:trigger_strategy}
\begin{figure}
	\centering	
	\includegraphics[width=1.00\textwidth]{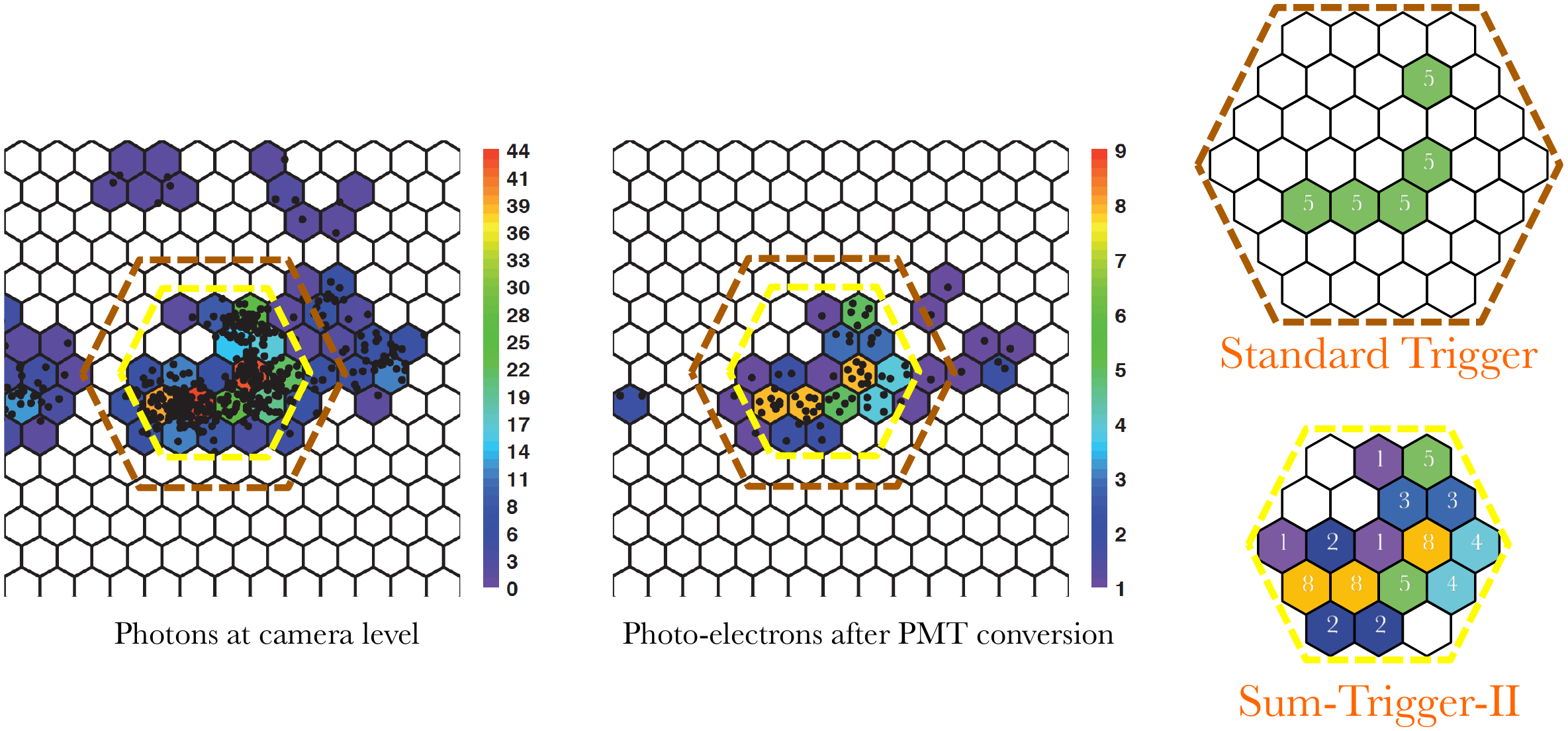}
	\caption{Simulated signal produced by a 31.8\,GeV gamma-ray shower in one of  the MAGIC cameras. On the left, the black dots represent the photons collected at the camera plane. In the centre, a huge part of these photons is lost due to the non perfect transmittance of the camera entrance window, reflectivity of the light concentrators and photon detection efficiency of the PMTs. Only around 20\% of the entire photon population is converted in photo-electrons and processed by the trigger system. On the right, comparison of the response of the standard trigger and Sum-Trigger-II to the same event.}
	\label{fig:low_energy_event}
\end{figure}

An air shower generated by a low energy gamma ray, of for instance 30\,GeV, develops for a short path in the atmosphere. The cascade dies out quickly and the resulting photon density at ground is very often weak, with a modest transversal development and not compact. Several photo-sensors are hit by Cherenkov photons, but only few of them collect enough photons that can be converted in considerable electrical signals of some photo-electrons. Respect to the digital trigger, that requests a certain amount of charge in every pixel belonging to a close-compact configuration, Sum-Trigger-II relaxes this constrain accepting any charge contribution. Figure \ref{fig:low_energy_event} shows the same event seen by the standard trigger and a module (hexagon of 19 photo-sensors) of Sum-Trigger-II.
The standard MAGIC system would not trigger because the close-compact requirement is not satisfied, even if five pixels exceed the typical threshold of around 5\,Ph.E. In Sum-Trigger-II, even pixels with a small signal well below the background level contribute to the final trigger decision. The final sum of all the 19 pixels returns a huge signal containing more than 50\,Ph.E. that is abundantly beyond the typical threshold of around 20\,Ph.E.. 

\subsection{System layout and stereoscopic implementation}
	\label{subsec:system_layout}
\begin{figure}[htb]
	\centering
  	\subfigure[Front view]
  	{ 				
		\includegraphics[width=0.388\textwidth]{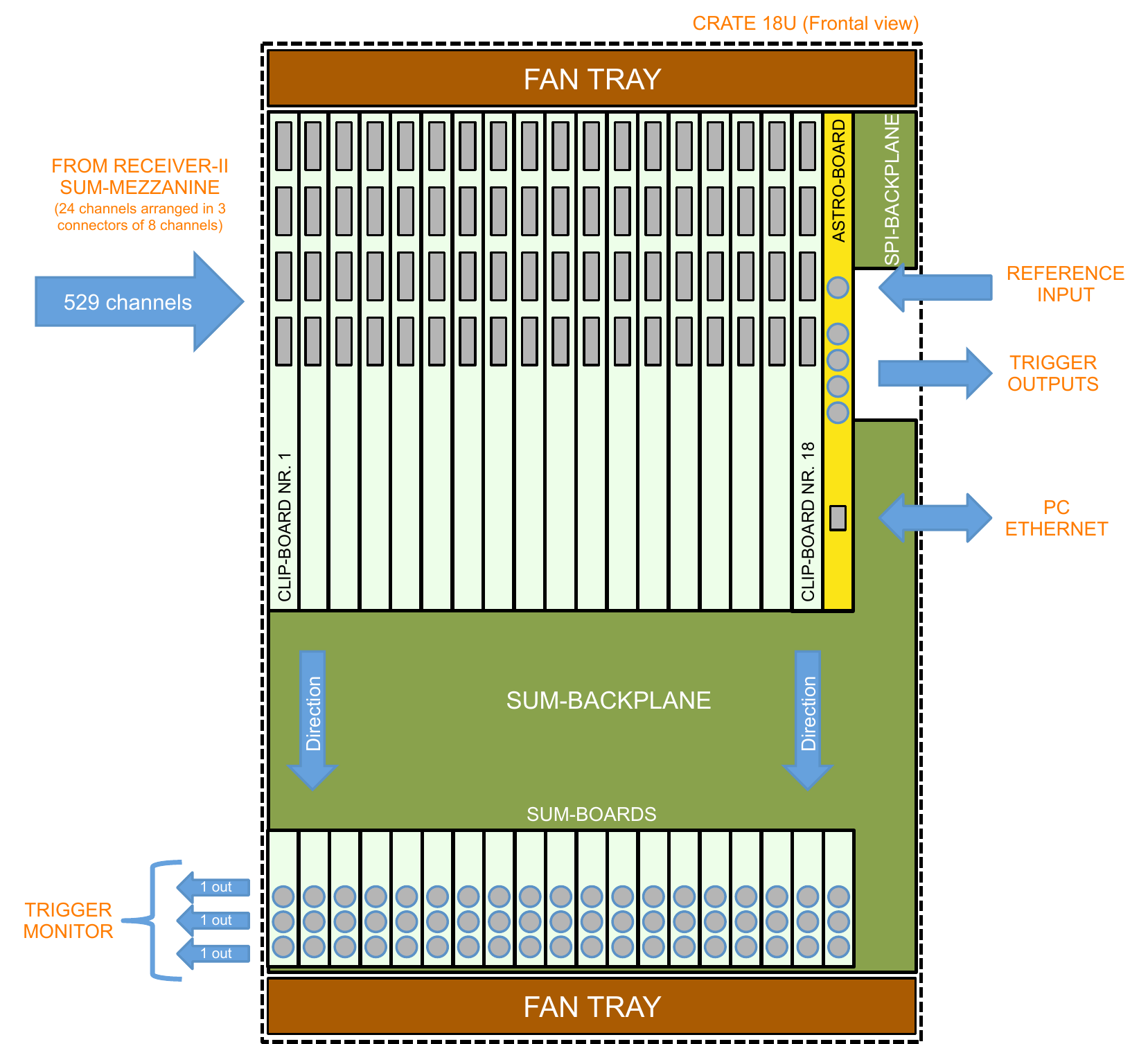}
		\label{fig:sketch_rack_front}
  	}	
  	\subfigure[Side view]
  	{ 				
		\includegraphics[width=0.335\textwidth]{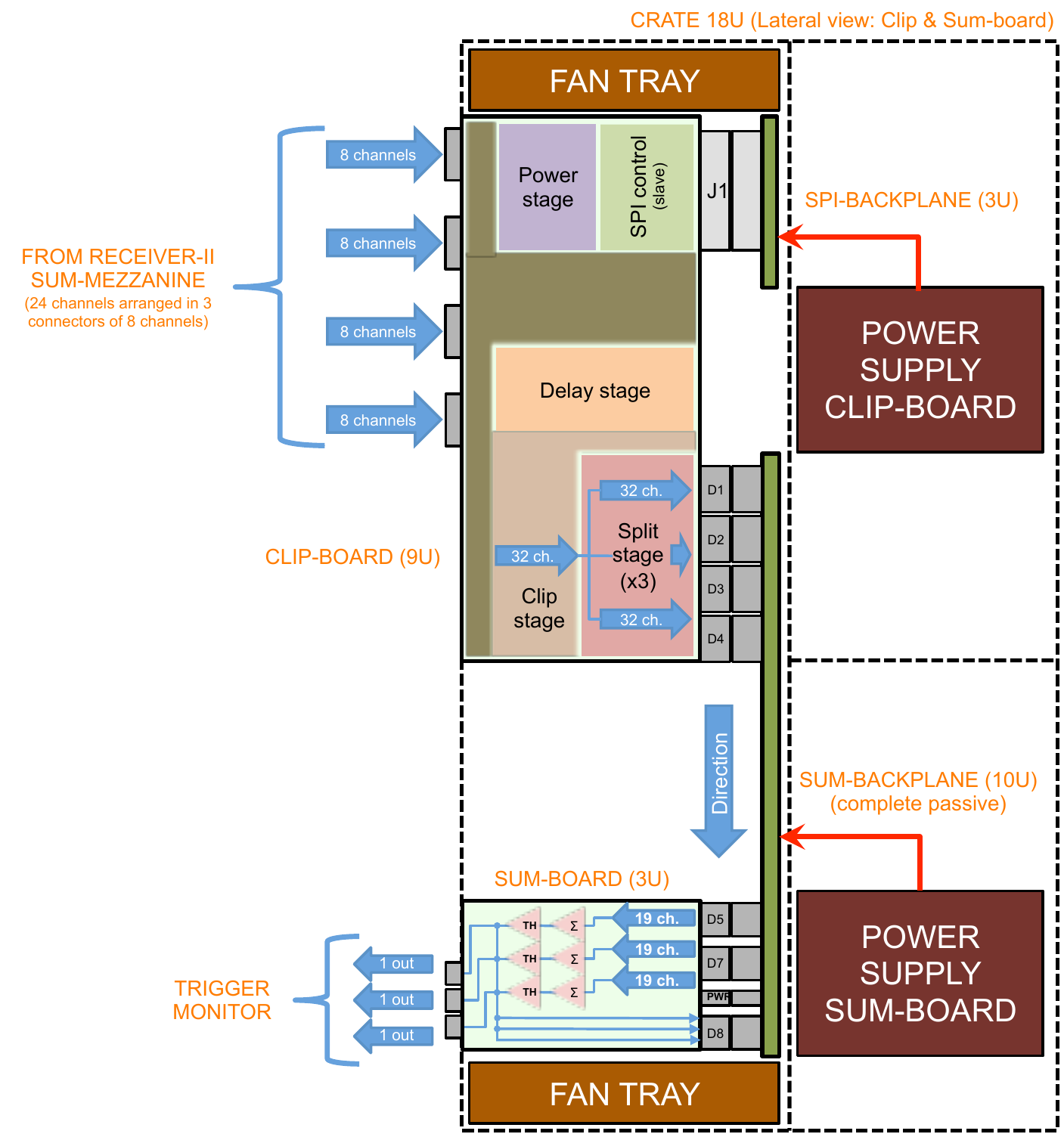}
		\label{fig:sketch_rack_side}
	}	
  	\subfigure[Photo]
  	{ 				
		\includegraphics[width=0.208\textwidth]{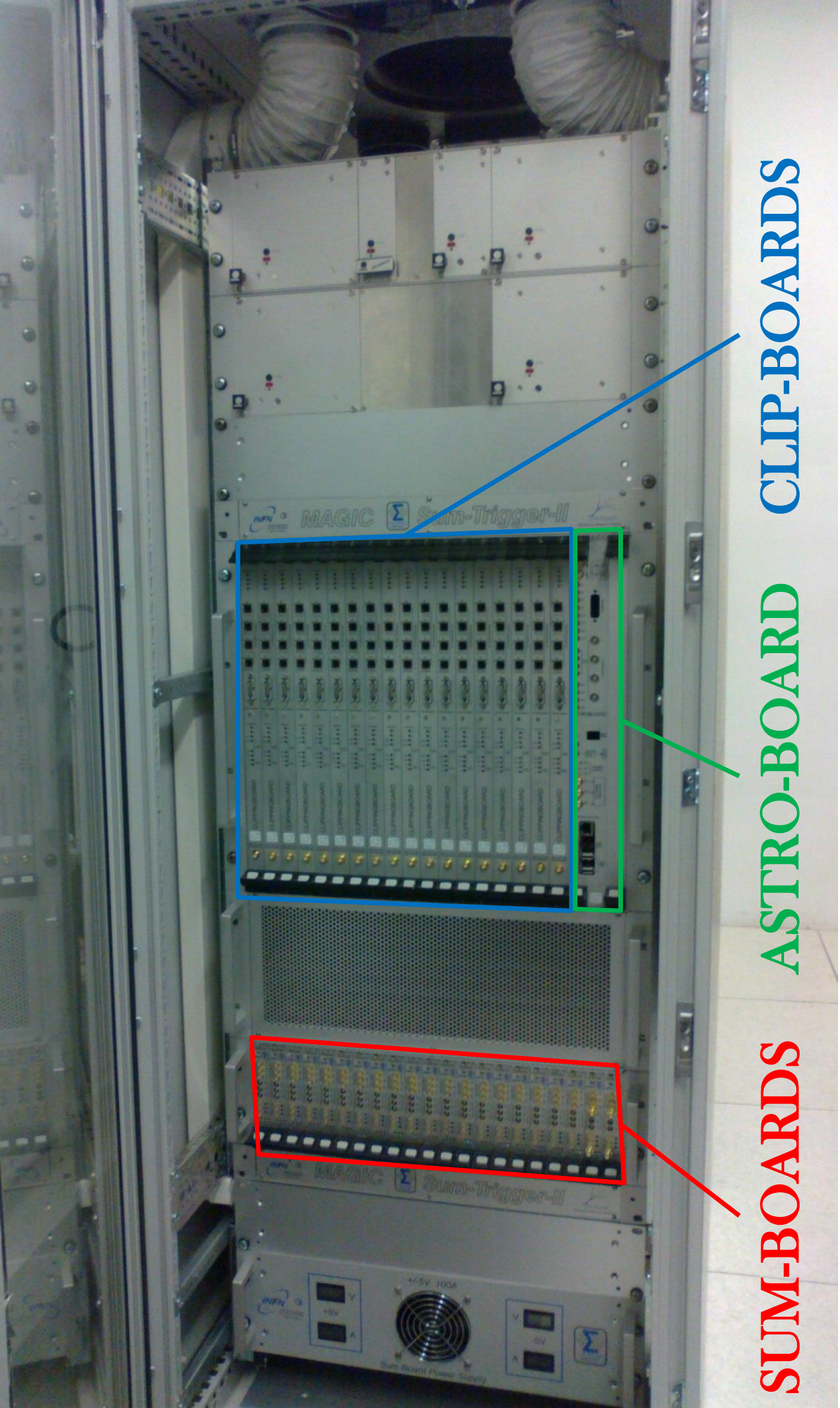}
		\label{fig:photo_rack}
	}	
	\caption{\textsl{a:} Front view of the simplified sketch of the Sum-Trigger-II rack. \textsl{b:} Side view of the simplified sketch of the Sum-Trigger-II rack. Delay, amplitude and clipping adjustments are perform in the Clip-boards, the summing stage in the Sum-boards, and the communication control in the Astro-board. \textsl{c:} Photo of the Sum-Trigger-II rack.}
	\label{fig:rack}
\end{figure}

The Sum-Trigger-II is a modular system covering 96.5\% of the standard trigger area, allowing data taking in wobble mode. This was not possible with the previous Sum-Trigger that had a ring-like layout instead of the full-circle one of this new design. Its area is divided into 55 partially overlapping sub-regions of 19 pixels, dubbed macrocells. Every macrocell works independently and its channels are piled-up after an accurate equalization in time and amplitude. These operations are essential for getting rid of the dis-homogeneity of the gain and propagation time of the electronics chain (photo-sensors included). It is worth to mention that before the summing stage (Sum-boards) the pulses are clipped (Clip-boards) in order to nullify the probability to trigger on single huge pulses created by the after-pulsing effect \cite{after-pulse}. A programmable discriminator threshold is applied just after the summing stage and a ``macrocell trigger'' is generated every time the signal exceeds the threshold for more than 700\,ps. In the final stage, all the ``macrocell trigger'' outputs are combined (OR-logic) in a single ``telescope trigger'' that it is used for the stereoscopic trigger and for reading out the event. The full Sum-Trigger-II flow chart has been already presented in other documents \cite{dazzi_phd} \cite{sum_icrc13}, while hereafter figure \ref{fig:rack} shows an explicative sketch and a photo of the Sum-Trigger-II system.

The entire system is controlled by a C++ multi-thread software, dubbed CRISTAL. It runs in an embedded PC plugged in the Astro-board that also hosts a powerful FPGA. CRISTAL sets the attenuations, delays and clipping levels of the Clip-boards, it changes the thresholds of the Sum-boards and it monitors the temperatures and the trigger rates. Finally, it communicates with the central control program of the telescope. 

\begin{figure}
	\centering	
	\includegraphics[width=0.90\textwidth]{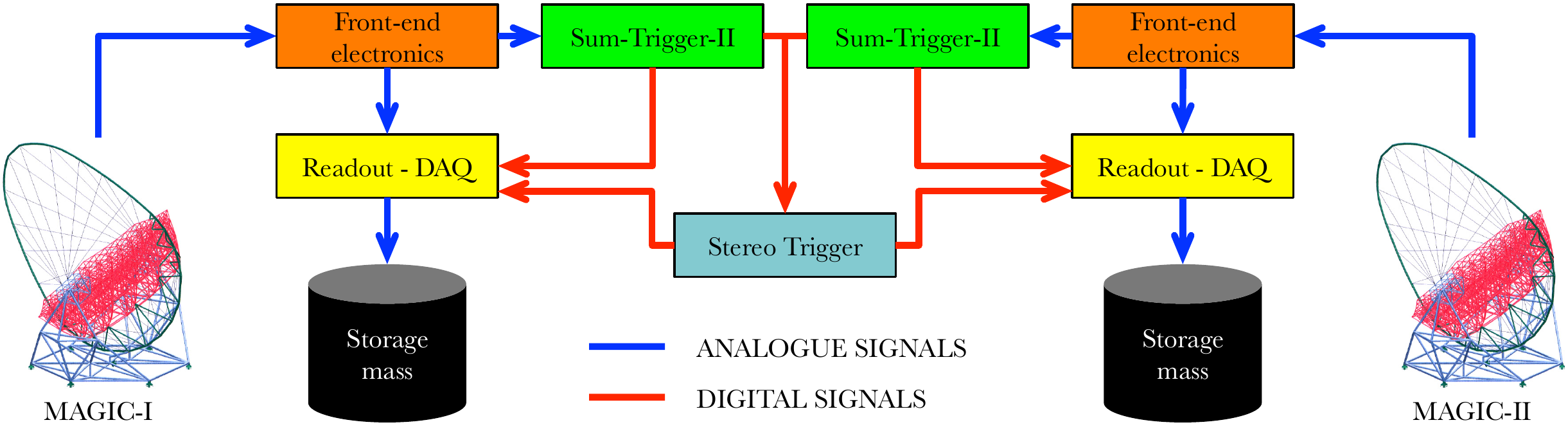}
	\caption{The stereoscopic Sum-Trigger-II scheme.}
	\label{fig:stereoscopic_scheme}
\end{figure}

One of the main innovations with respect to the prototype analogue trigger constructed in 2008 is the stereoscopic implementation. Both MAGIC telescopes are equipped with a Sum-Trigger-II system and can be used in stereo data-taking mode. Figure \ref{fig:stereoscopic_scheme} shows the stereoscopic scheme.
The enlarged trigger area assures a sufficient overlap between the trigger fields of view of the telescopes, that strongly affect the collection area of the instrument. The constrain to ask for coincident\footnote{The acceptance time gate is set to 180\,ns.} events in both telescopes drastically reduces the accidental triggers and allows to lower the discriminator thresholds without saturating the readout system. Another advantage is that the stereoscopic implementation partially does the work of the clipping stage. The probability that huge after-pulses occur at the same time in both telescopes is reduced by two orders of magnitude. In this way it is possible to apply soft clipping cuts and consequently reduce the amount of rejected gammas.

\section{Performances}
	\label{sec:performances}
The concept of the Sum-Trigger-II is simple, but a perfect tuning of some key parameters is crucial. An effective pile-up occurs only when all the pulses are aligned on time and the charge contributions are not affected by differences in the gain. This approach increases the signal to noise ratio, but with the assumption that the noise is uncorrelated. Correlated noise present in several neighbouring channels is easily accumulated and it can generate fake triggers. The signal integrity plays also an essential role in this strategy. The most critical point of the system is the clipping stage. This signal processing must not introduce any artificial contribution and it must behave in the same way in all the channels. Summarizing the key parameters for a functioning analogue trigger are: precise time and gain adjustment, stable clipping stage and low coherent noise.

\subsection{Calibration}
	\label{subsec:calibration}
\begin{figure}[htb]
	\centering
  	\subfigure[Amplitude distribution]
  	{ 				
		\includegraphics[width=0.49\textwidth]{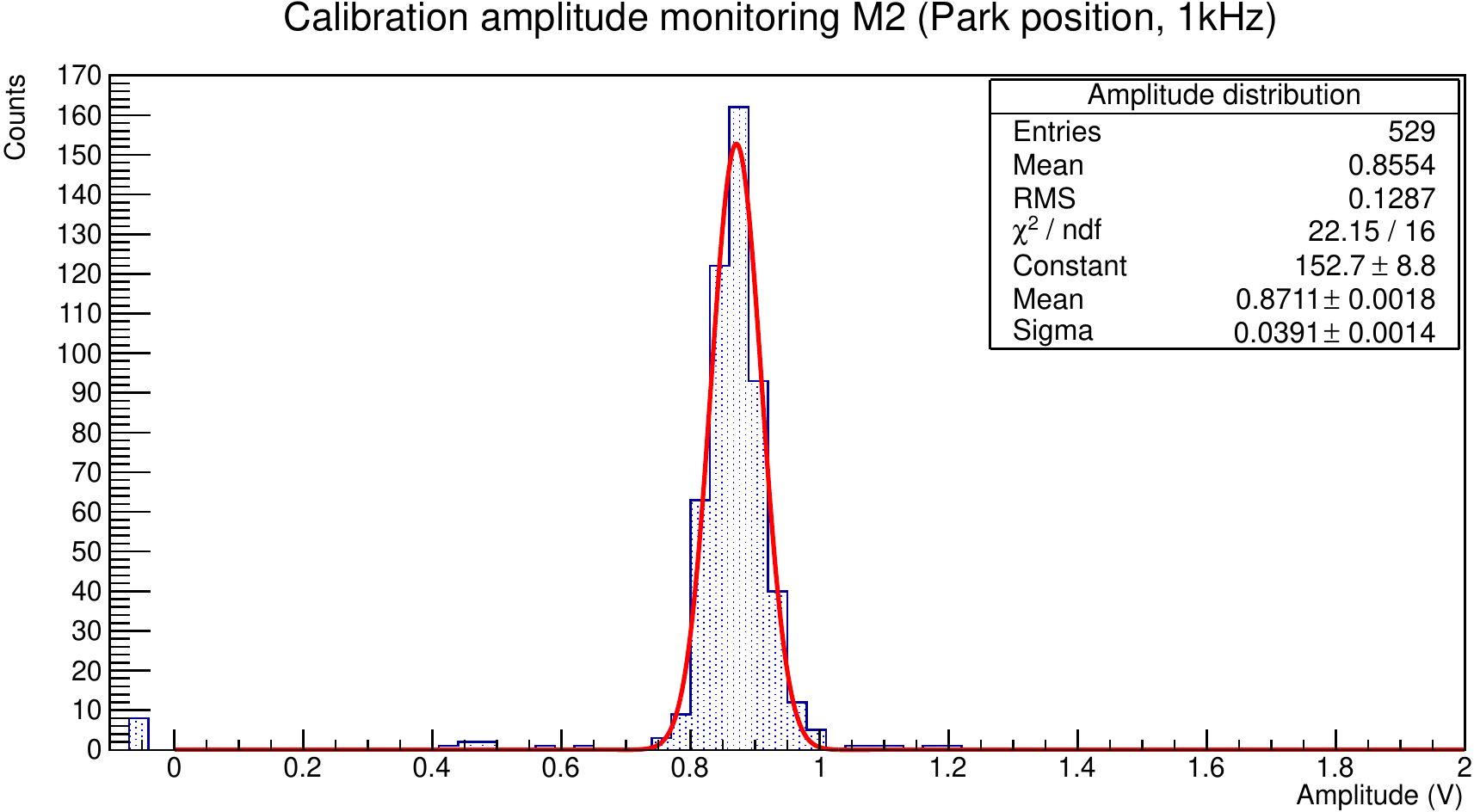}
		\label{fig:after_amplitude}
	}	
	\hspace{0.05\textwidth}
  	\subfigure[Calibrated waveforms]
  	{ 				
		\includegraphics[width=0.40\textwidth]{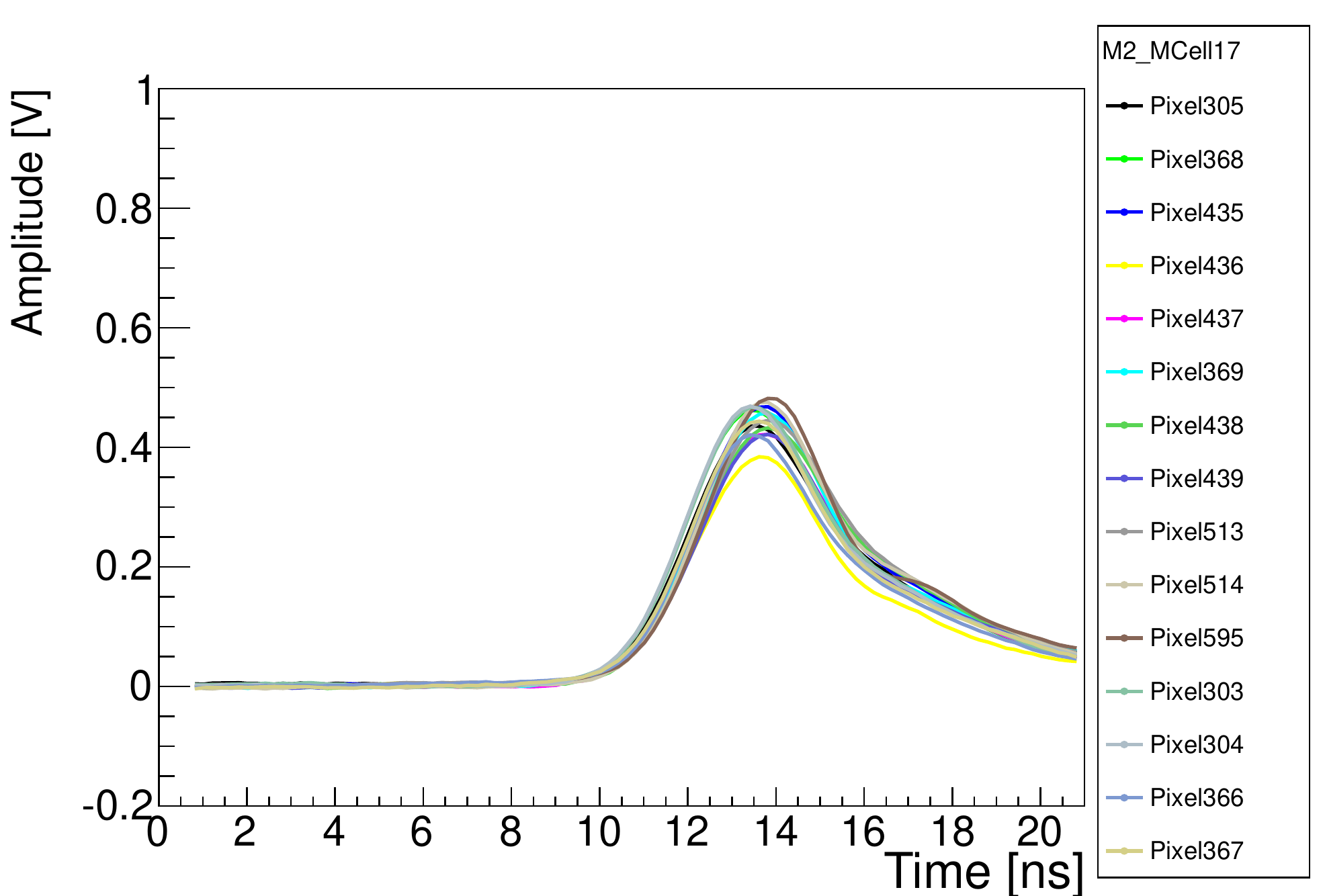}
		\label{fig:calibrated_pulses}
	}	
	\caption{\textsl{a:} Pulses amplitude distribution of all the channels belonging to the trigger area after the amplitude calibration process. \textsl{b:} Waveforms of the 19 channels belonging to a single macrocell after a single step calibration process (amplitude and time).}
	\label{fig:calibration}
\end{figure}

The calibration is an iterative procedure based on delay and threshold scans that sequentially adjusts the amplitude and the propagation time of the pulses before the summing stage. It makes use of attenuators with a resolution of 0.5\,dB over a dynamic range of 31.5\,dB and continuous programmable delay lines mounted on the Clip-boards. After every cycle, a distribution of pulse's delays and amplitudes is generated. Then, new parameters for the delay lines and attenuators are computed in order to approach the mean of the distribution Gaussian fit. Figure \ref{fig:calibration} shows the pulses amplitude distribution after the adjustment and how the real pulses, recorded with the oscilloscope, look like when also the delay correction is applied.
The amplitude dispersion is $\sim$\,5\%, which is close to the limit imposed by the attenuator resolution. The delay dispersion is $\sim$\,200\,ps, but it can be improved as explained in section \ref{sec:future_improvements}. The clamp circuitry has been evaluated by fixing the same clipping level in all the pixels and then measuring the amplitudes running a threshold scan. The outcome demonstrates that the clipping level is equally precise in all the channels within the resolution of the adopted method (< 5\%). 

\subsection{Noise}
	\label{subsec:noise}
\begin{figure}[htb]
	\centering
  	\subfigure[Sub-systems noise scan]
  	{ 				
		\includegraphics[width=0.44\textwidth]{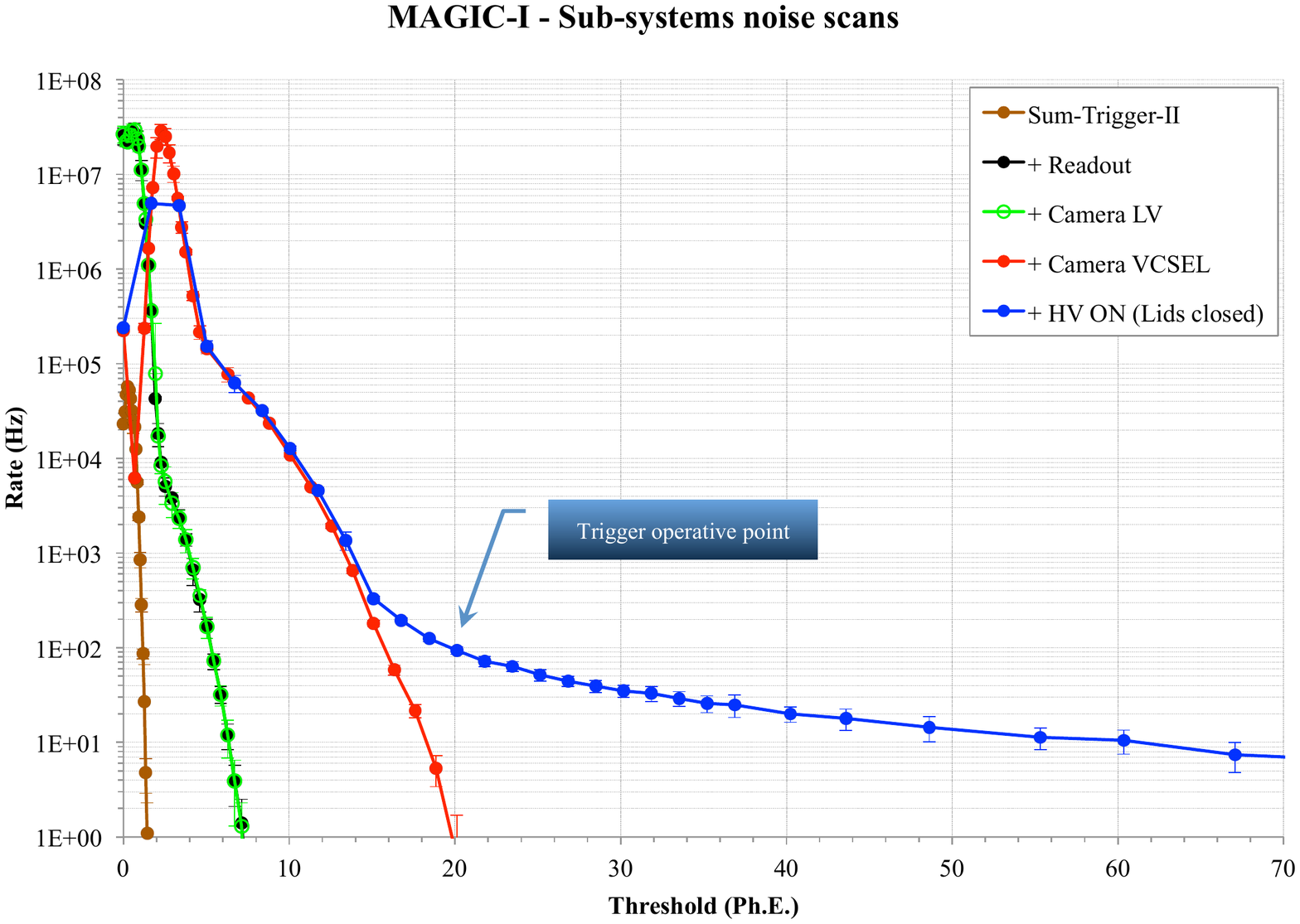}
		\label{fig:sub-system_noise}
  	}	
  	\hspace{0.06\textwidth}
  	\subfigure[Sub-regions noise scan]
  	{ 				
		\includegraphics[width=0.44\textwidth]{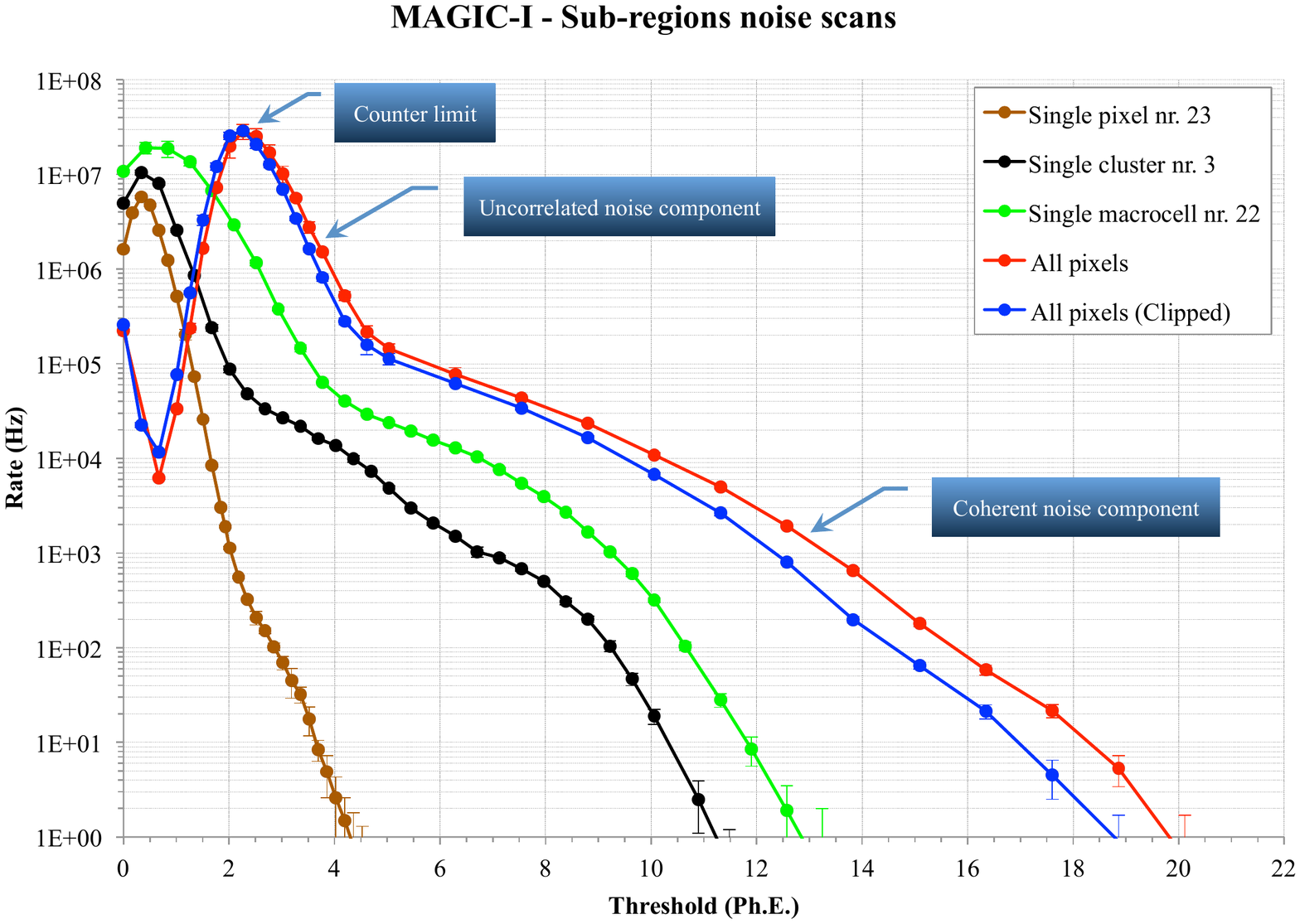}
		\label{fig:sub-regions_noise}
	}	
	\caption{\textsl{a:} Noise rate scans by sequentially turning on the sub-systems that belong to the MAGIC electronics chain. \textsl{b:} Noise rate scans of different sub-regions of photo-sensor camera.}
	\label{fig:noise_scans}
\end{figure}

Two types of noise rate scans have been performed by measuring the trigger rate as a function of the threshold. The first one is focused on decoupling the noise contributions generated by every sub-system of the electronics chain. The noise level has been measured with the camera lids closed and by switching on the sub-systems in the following order: 1) Sum-Trigger-II; 2) Receiver boards\footnote{Boards that receive the optical signal from the camera and split it into the trigger and readout branch.}; 3) Camera low voltage; 4) VCSEL transmitters\footnote{Transmitters placed inside the camera that send the photo-sensor signals to the Receivers over long optical fibres.}; 5) PMTs high voltage. Figure \ref{fig:sub-system_noise} demonstrates that the contribution of Sum-Trigger-II is less than 2\,Ph.E. (sum of 19 channels) and it is negligible compared to the rest of the electronics chain. The second noise rate scan (Figure \ref{fig:sub-regions_noise}) has been performed with the camera lids closed, the PMTs high voltage off and the electronics chain on. Five different configurations have been tested by activating the VCSEL (signal transmission) of a single pixel, a single cluster (7 PMTs), a single macrocell and the full camera with and without clipping. The trend of the noise enhancement in the five configurations reveals that the noise has two components, whose one is coherent. However, its contribution is negligible at the operative point around 20\,Ph.E.. 

The noise rate scan of the entire camera has been repeated in different periods showing a very good reproducibility. The same study has been done on MAGIC-II, showing similar performances of trigger system. However, the test has revealed the presence of too high coherent noise outside Sum-Trigger-II that it is under investigation.

\subsection{Rate scans}
	\label{subsec:rate_scans}
\begin{figure}
	\centering	
	\includegraphics[width=0.80\textwidth]{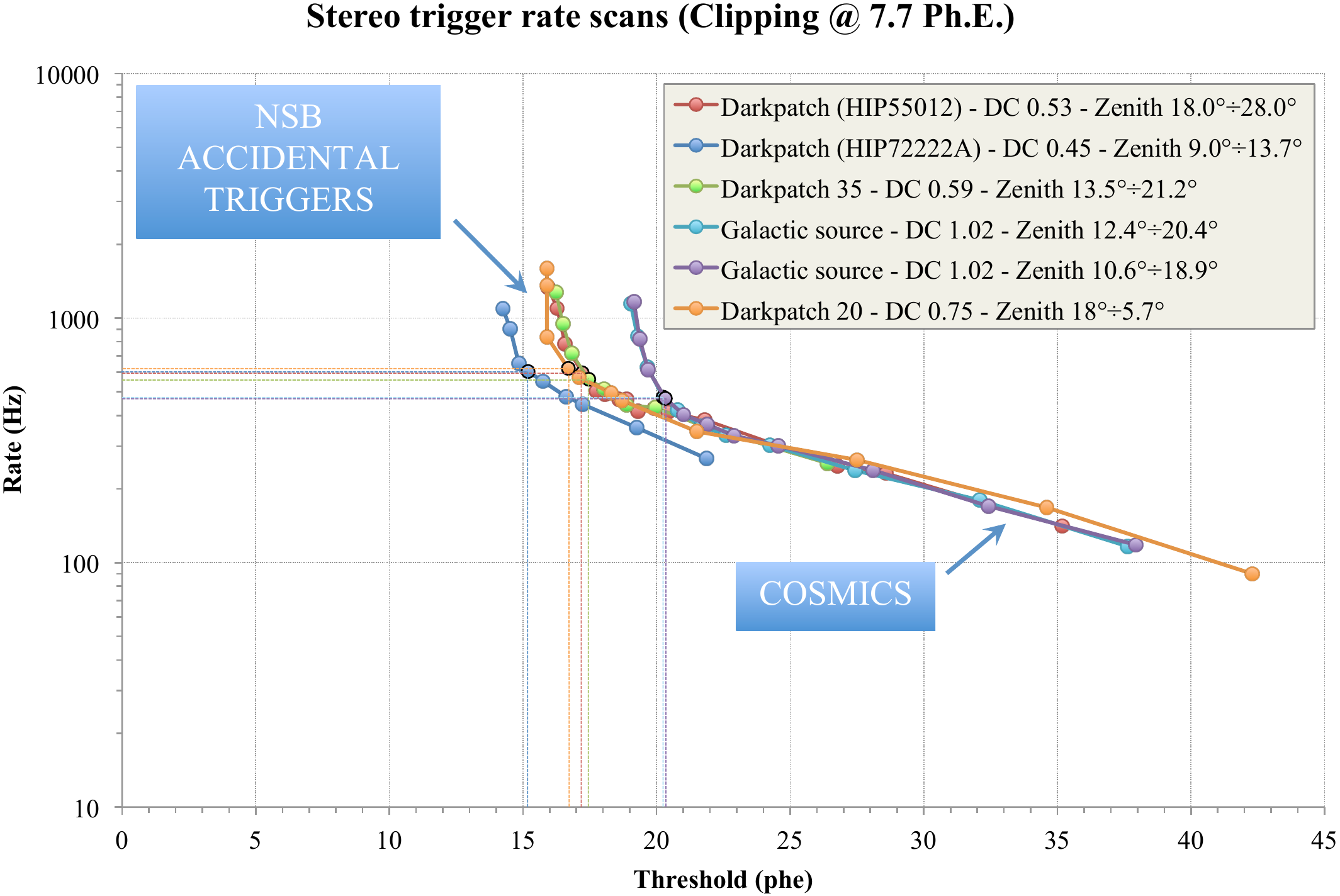}
	\caption{Total trigger rate as a function of the discriminator threshold for different field of views like galactic sources or extra-galactic dark-patches. Every curve has two main contributions: on the left the steep slop mainly due to the Night Sky Background and on the right the one due to cosmic ray events.}
	\label{fig:trigger_rate_scan}
\end{figure}

Several stereoscopic trigger rate scans as a function of the discriminators threshold applied to the summed signals have been performed during good weather conditions. Different light conditions and zenith angles lower than 30 degrees have been selected for these scans. Some of them have been repeated with very similar conditions demonstrating that the measurements are reproducible and the threshold at the working point easily deducible. The most interesting rate scans are presented in figure \ref{fig:trigger_rate_scan}.
The working points are marked by the intersection of the coloured segments. The threshold is $\sim$\,20\,Ph.E. for galactic sources and $\sim$\,17\,Ph.E. for extra-galactic ones. The typical stereoscopic rate is 500 - 600\,Hz by fixing the single telescope rate to 30\,kHz. 160\,Hz out of 500 - 600\,Hz are still accidental triggers that need to be eliminated off-line. Considering that the average NSB rate is of the order of 100\,MHz, the rejection factor is between 10$^5$ to 10$^6$ at the working point.

\section{Future improvements}
	\label{sec:future_improvements}
Due to the fact that the standard trigger area is slightly larger than Sum-Trigger-II one, some high energy events with high impact parameter could be lost. This minor inefficiency could be solved by merging both trigger systems. In addition, Monte Carlo studies reveal that the trigger energy threshold can be lowered by combining Sum-Trigger-II with a relaxed clipping level and the standard trigger looking for 2NN patterns.

The time adjustment of the pulses is crucial for the efficiency of the trigger. The current default version of the time calibration routine spans from 0\,ns to 6.8\,ns with a sampling of 200\,ps and a statistics of 1000 events. The time resolution of the programmable delay lines and the desired statistics can be arbitrarily selected. However, higher is the time resolution and the statistics, and longer is the time needed to perform the time calibration process. It needs to be studied which is the best combination of sampling and event statistics by keeping the same calibration process duration.

The MAGIC stereo trigger has been recently equipped with a new sub-system, dubbed Topo-trigger, that adopts topological constrains for rejecting accidental events \cite{topo}. Currently, it can only accept information from the macrocells of the standard trigger, but with a proper interface it can accommodate Sum-Trigger-II too. Compared to the standard trigger area that is covered by 19 big macrocells, Sum-Trigger-II counts 55 small macrocells. This can assure to Topo-trigger a better spatial resolution that increases the rejection power. 

\section{Conclusions}
	\label{sec:conclusions}
Gamma-ray observations are still limited by the background at lower energies. A better background suppression through effective trigger strategies is essential for covering the energy domain in overlap with satellite experiments.	Sum-trigger-II is an alternative to the current standard digital trigger, which has the capability to lower the energy threshold of the MAGIC telescopes. The technical performances fulfil the expected requirements and there is still room for improvements. Finally, it is kept open the possibility to combine both trigger systems or to implement the Sum-Trigger-II into the stereoscopic Topo-trigger for a further step forward.

\acknowledgments
The authors would like to thank the Instituto de Astrofisica de Canarias for the excellent conditions at the Observatory del Roque de los Muchacos in La Palma and Altera Inc. for supporting the project through the Altera University Program. The support of the German BMBF and MPG, the Italian INFN and the Spanish MICINN is gratefully acknowledged. Several colleagues have supported the technical activities of the project. It is worth to mention: D. Corti, A. Dettlaff, D. Fink, J.L. Herrera, R. Maier, C. Manea, T. S. Tran and M. Will.



\end{document}